\documentstyle[aas2pp4]{article}

\begin{document}

\title{ On the kinematic signature of a central Galactic bar in observed star
samples}

\author{P.~Vauterin}
\affil{Royal Observatory of Belgium, Ringlaan 3, B-1180 Brussel, Belgium}
\author{H.~Dejonghe}
\affil{Universiteit Gent, Sterrenkundig Observatorium, Krijgslaan 281(S9),
       B--9000 Gent, Belgium}

\begin{abstract}
A quasi self--consistent model for a barred structure in the central regions of
our Galaxy is used to calculate the signature of such a triaxial structure on
the kinematical properties of star samples. We argue that, due to the presence
of a velocity dispersion, such effects are much harder to detect in the stellar
component than in the gas. It might be almost impossible to detect stellar
kinematical evidence for a bar using only $l$-$v$ diagrams, if there is no a
priori knowledge of the potential. Therefore, we propose some test parameters
that can easily be applied to observed star samples, and that also incorporate
distances or proper motions. We discus the diagnostic power of these tests as a
function of the sample size and the bar strength. We conclude that about 1000
stars would be necessary to diagnose triaxiality with some statistical
confidence.
\end{abstract}


\section{Introduction}

During the last years, there has been mounting evidence that the central part of
our Milky Way contains a triaxial, barlike structure. This theory is supported
by the observation of the ``parallellogram'' structure in the kinematics of the
gas (see e.g. \cite{Binney91}), and by the asymmetry in both the integrated
light (e.g. \cite{Blitz91}) and the star samples (e.g. \cite{Whitel92}).

However, as pointed out already by de Zeeuw (\cite{Zeeuw94}), it is striking
that there is currently almost no evidence for such a bar in the stellar
kinematics.  The only observations which seem to point in that direction are
the vertex deviations in the samples of K-giants (\cite{Zhao94a}), but these
data suffer from large statistical uncertainties. Put differently, triaxial and
axisymmetric models fit almost equally well kinematical data (\cite{Blum95}).
Of course, this is an unsatisfactory situation, because the stellar component
is the most massive one and thus should play a dominant role in the creation
and evolution of such a bar.

Therefore, we will address two questions in this paper: (1) which are
the dominant effects caused by a triaxial structure in the kinematics
of stellar samples, and (2) what is the magnitude of these effects. In
this way, we will try to determine where the triaxiality shows up, as
well as the required minimal sample size and observational accuracy in
order to be able to detect it.

In order to achieve this goal, we use a self--consistent model for the stellar
component of a barred galaxy which we developed in an earlier paper (\cite{Vau97}).
The various scale parameters are adjusted in such a way that
the model is a fair fit to the central region of our Milky Way.  Since the
distribution function is known with high accuracy, it is easy to draw synthetic
stellar samples from this model. 

We discuss the effect of the bar on the $l$-$v$ diagram,
and further construct various test parameters for triaxiality, based on the
kinematical properties of individual stars in stellar samples. These tests
involve the radial velocities, distances and/or proper motions. Using Monte
Carlo simulations, we calculate the distribution of these parameters for
different sample sizes, in order to determine minimal sample sizes.

\section{A model for the stellar component of a bar in the centre
of the Galaxy}

We apply a model which we constructed in an earlier paper (\cite{Vau97};
hereafter paper I). It is based on the nonlinear extension of
an unstable linear mode occurring in an initially axisymmetric,
two--dimensional model. The galaxy consist of two components:

\begin{itemize}
\item
An unperturbed, axisymmetric and time--independent part with a binding
potential $V_0$ and a distribution
function $f_0$. The parameters (scale and rotation curve) are
adjusted in order to fit the central region of the Galaxy. The
rotation curve of this model is displayed in fig. \ref{fig_axi}, and 
additional details about this model can be found in paper I.
\item
A barlike perturbation, which is non--axisymmetric and time--dependent. The
binding potential is denoted by $V'$, and the distribution function is $f'$.
The time and angular dependency of this part are of the following form:
\begin{equation}
  \Re\left[ e^{i(m \theta - \omega t)} \right],
\end{equation}
where $m$ is the symmetry number, $\Re(\omega)/m$ the rotation speed and
$\Im(\omega)$ the growth rate.
\end{itemize}

\placefigure{fig_axi}

The motion of the stars in the perturbation is a solution of the Boltzmann
equation:
\begin{equation}
{\partial f' \over \partial t}-[f_0',V']=[f_0,V']+[f',V'].
\end{equation}
In addition, for a self--consistent perturbation, the Poisson equation
applies:
\begin{equation}
\nabla V'= 4 \pi G \rho'.
\end{equation}

When the perturbation is sufficiently small, one can linearize the equations.
In this case, the last, quadratic term of the Boltzmann equation is omitted.
Solutions ($V'$,$f'$) that satisfy both the Poisson equation and the
linearized Boltzmann equation are called linear modes. There exist various
methods in the literature to solve this problem (e.g. \cite{Kaln77},
\cite{Hunter92}, \cite{Vau96}). The analyses show that, for each value of
$m$, an infinite series of linear modes exists, with different rotation speeds
and growth rates. The mode with the highest growth rate is the dominant
instability. In most cases, it turns out to be an $m=2$ barlike structure (as
is the case for the present model).

By construction, linear modes are only valid for infinitesimally small
amplitudes of the perturbation. In order to construct finite amplitude
perturbations, we performed a nonlinear extension. The total potential is taken
from the linear mode, $V_{\rm TOT}= V_0 + \epsilon V'$, but the corresponding
distribution function is obtained by solving the full, nonlinear Boltzmann
equation numerically. The details about these calculations are given in paper I.

The  total distribution function of such a barred model can be accurately
calculated for any point in phase space. In addition, it is possible to
calculate derived quantities, such as mass densities, streaming velocities, 
etc., by numerical integration over a grid in the velocity coordinates
\label{modeldef}

In this way, we constructed a physically meaningful barred model, which is
self-consistent to a large degree. The bar has a semi--major axis of about 2
kpc, a semi--minor axis of about 1 kpc, and has a 15 degrees tilt angle
with respect to the direction of the sun. In addition, we have put the sun at a
distance of 8 kpc from the center of the Galaxy. These values seem to be more
or less compatible with most of the observations (see e.g. \cite{Zhao94}).

\section{Stellar $l$--$v$ diagrams of the bar}

\label{sect_lv}

In the top panels of fig. \ref{fig_lv}, we compare the stellar $l$-$v$ diagram
of the barred model to the one of the unperturbed system. Apart from density
changes, the presence of the bar also clearly  has some effects on the
kinematical properties of the $l$--$v$ diagram. The most important difference
is that the average radial velocity of the stars is higher than the rotation
curve of the axisymmetric potential. This is easily explained by the fact that
we see the bar almost along the major axis, and that the stars crossing the
minor axis are moving faster than those crossing the major axis of the bar (see
e.g. fig. 7 of paper I). The same figure also shows the $l$--$v$ diagram of the
barred model, observed from different points. The presence of a bar also clearly
introduces substantial changes in the structure of these diagrams.

\placefigure{fig_lv}

However, if the potential is not known a priori, it is not possible to easily
conclude that these $l$--$v$ diagrams are caused by a barred system. They could
very well be compatible with an axisymmetric system that has a steeper or a
slower rotation curve. Therefore, the $l$-$v$ diagram alone is usually not
sufficient to prove triaxiality, and one needs to incorporate other information
as well. This is in contrast to the situation for the gas, where the bar
introduces orbits that are ``forbidden'' by an axisymmetric potential.
Unfortunately, there exist no such forbidden regions for stars, due to the
presence of a velocity dispersion.

In addition, the features induced by triaxiality are much less pronounced in
the stellar $l$--$v$ diagram than in the gaseous counterpart. Again, this is
mainly due to the important velocity dispersion of the stars, which tends to
smooth out the various properties of the different orbital families (see also
paper I). The situation for the gas is much simpler, since, in a good
approximation, the behaviour of the system at each point is completely
determined by the single, closed orbit passing through that point.

\section{Tests including additional information}

As mentioned in the previous section, the galactic length $l$ and the radial
velocity $v_r$ of the stars in a sample are not sufficient to make a clear
distinction between a triaxial and an axisymmetric system. Therefore, we will
investigate two additional types of information: distances and proper motions.
It is well known though that such information, if present at all, is subject to
rather large errors. A test should therefore not rely too critically on the
exact values of these quantities, but should be rather robust to errors.

\placefigure{fig_sceme}

In fig. \ref{fig_sceme}, the streaming velocities of the stars in a barred
system are shown in an exaggerated and schematic way. The streamlines have an
elliptical form, aligned along the major axis of the system. In addition, the
mean velocity is larger for stars that cross the minor axis than for
those on the major axis. It is clear that tests for triaxiality should take
advantage of these properties in some way.

The test parameters presented in the following sections are functions of the
observable parameters of a star sample taken from the centre of our Galaxy.
Therefore, we need simulated star samples,  in order to be able
to discuss the properties of these parameters. One can easily draw such a star
sample from the stellar distribution function using a try--and--reject
technique. In this approach, a random number generator creates star
coordinates (positions $\vec r$ and velocities $\vec v$, both in rectangular
reference systems) that have a rectangular distribution over the whole area of
interest. For each star, an additional random number $R$ is further generated,
having a rectangular distribution between 0 and the maximum value reached by
distribution function (for our models, this corresponds to the central value).
The coordinates ($\vec r,\vec v$) are accepted as a new member of the star
sample only if the value of $R$ is smaller that the value of the distribution
function at that point; it is rejected in the other case. This process is
continued until the sample contains the desired number of stars.

\subsection{Distance information and radial velocities}

As shown in fig. \ref{fig_sceme}, the line of zero mean line--of--sight (LOS)
velocity in a barred  system with the adopted orientation is in general not
aligned anymore with the line  $l=0$ (as is the case for an axisymmetric system).
As a consequence, there is a symmetry breakdown in the $l$-$v$ diagram: the
part of the diagram that contains stars that are closer to the sun than the
galactic centre is not identical anymore to the part that contains the stars
lying farther away. 

\placefigure{fig_distlv}

In this section, we will exploit this asymmetry, and use it as a basis for a
test parameter for the triaxiality of the system. To this end, we subdivide the
stars into eight different groups, labelled $(xxx)$, where $x$ can be $-$ or
$+$ (see also fig. \ref{fig_distlv}). The first sign indicates the distance from
the sun ( $-$ means closer than the galactic centre and $+$ means further
away), the second sign is related to the galactic longitude ($-$ for negative and
$+$ for positive longitudes), and the last sign is determined by the radial
velocity ($-$ for negative values and $+$ for positive values).

The number of stars $N_{xxx}$ in each subclass is used in order to construct a
test parameter:

\begin{equation}
T_1= {1 \over 2} \left( {
      N_{--+} \ /\ N_{---}
\over N_{-+-} \ /\ N_{-++} } 
+ 
{     N_{++-} \ /\ N_{+++} 
\over N_{+-+} \ /\ N_{+--} 
} \right) \times 100.
\end{equation} 

The denominators of both fractions contain the fraction of counterrotating stars in
the minor
axis quadrants, at the near side (first fraction), and at the far side (second
fraction). On the other hand, the numerators contain the fraction of counterrotating
stars in
the major axis quadrants in the corresponding distances classes. Therefore, this
parameter
essentially expresses the fact that the quadrants containing the major axis of the bar
(i.e.
$N_{-+x}$ and $N_{+-x}$) contain more counterrotating stars than the other quadrants
($N_{++x}$ and $N_{--x}$).

Because we take only ratios of stars in one give distance class, the
parameter is insensitive to photometric effects (caused by the fact that there is a
bias
against seeing stars at the far side of the bar), and is thus mostly determined
by the kinematical properties of the sample. It only requires crude distance
information,
which is a prerequisite condition since distances are usually subject to large errors.

It is important to notice that large numbers of stars are summed in each
subclass, resulting in reduced noise effects. In addition, it is a so--called
"robust" test parameter, because only numbers of stars are involved instead of
measured values. Such a parameter is relatively insensitive to outliers.

\subsection{Proper motions}

A second test only involves proper motions. It is based on the fact that, in a
barred system which is not aligned with the line of sight, the mean proper
motions for positive and negative galactic longitudes have opposite sign (see also
fig. \ref{fig_sceme} and fig. \ref{fig_prop0}). This asymmetry is quantified
using a second test parameter, which measures the difference in proper motion
(PM) at positive an negative galactic longitudes.
\begin{equation}
T_2=\langle
PM_{l>0} \rangle - \langle PM_{l<0} \rangle.
\end{equation}

Unfortunately, this parameter is not robust with respect to errors in the
observed proper motions, so it might be less useful in practical circumstances.

\placefigure{fig_prop0}

\subsection{Proper motions and radial velocities}

It also follows from fig. \ref{fig_sceme} that, in a barred system that is not
aligned with the direction of the sun, the line of zero LOS velocity is not
equal to the $l=0$ line  (as is the case for axisymmetric systems). This
phenomenon causes differences in the partial $l$-$v$ diagrams containing stars
with respectively only positive and negative proper motions. In order to
quantify this difference, two subset are drawn from the star sample: (1) stars
having positive proper motions and lying in the upper $25\%$ of the proper
motion distribution, and (2) stars having negative proper motions and lying in
the lowest $25\%$. We rejected the middle $50\%$, because this is the ``gray
area'', where large measurement errors cause uncertainties on the sign of the
proper motion. Fig. \ref{fig_prop} displays the $l$-$v$ diagrams of both
subsets, and shows that the first subset has, on average, larger radial
velocities than the second. This difference is measured by the third test
parameter, defined as the difference between the mean radial velocities of both
subsets:

\begin{equation}
T_3= \langle v_{r,\ pm>75\%} \rangle - \langle v_{r,\ pm<25\%} \rangle.
\end{equation}

Again, this test parameter turns out to be fairly insensitive to photometric
effects. It is also robust with respect to errors on the measurement of proper
motions, since it depends only on their sign rather than on their exact values.
The fact that it is not robust with respect to the radial velocities is not a
disadvantage, because these values can be measured with rather high  accuracy.

\placefigure{fig_prop}

As one can infer from fig. \ref{fig_sceme}, the symmetry breakdown of the
radial motions and the proper motions turn out to work in the same direction
for the value of $T_3$. Obviously, this effect has a positive influence
on the discriminating power of this parameter.

\subsection{Monte Carlo simulations of the test parameters}
We used a Monte-Carlo simulation technique to numerically calculate the
statistical distributions of the test parameters for a given galaxy model.
A large set of random star samples, consistent with the
distribution function of the model, is constructed, and the values of
the various test parameters are calculated for each individual sample. 
The parameter values of all samples are further binned into a
histogram in order to calculate a numerical estimate of the distributions
of these parameters. Obviously, one should incorporate a sufficient
number of synthetical samples in order to obtain a  histogram with enough
resolution. We use histograms with 10 intervals, and a simulation set
containing 100 samples for each model. We have checked that this number is
sufficient to obtain a reasonable degree of accuracy by checking the
consistency of the results with estimations based on larger simulation sets.

\subsection{Probability distribution functions of the test parameters}
We calculated the distribution function of the parameters $T_1$, $T_2$ and
$T_3$ for two different galaxy models: an axisymmetric one and a barred system
with parameters adjusted to those of our Milky Way. These
distribution functions were calculated for samples containing 700 and 1400
stars. The resulting distributions are shown in fig. \ref{fig_pardst}.

\placefigure{fig_pardst}

\subsubsection{Verification of a hypothesis using observed data}

The combination of part [2] and [3] of the histogram (part [2] is the overlap
between both distributions) gives the  probability
distribution for the parameters in the case of an axisymmetric model. If one
has calculated a test parameter  using actual observations of our Milky Way,
the corresponding distribution can be used to check whether the value is
consistent with an axisymmetric model or not, given a specified level
of confidence.

On the other hand, part [1] and [2] correspond to the probability distribution
for the parameters of the barred model. Again, when actual observations are
present, this curve can be used to check the consistency with this model.

\subsubsection{Determination of the resolving power of a test parameter}

One can use the information in fig. \ref{fig_pardst} to estimate the
diagnostic power of a particular test for a given sample size. Let assume that,
in reality, the centre of our Milky Way is barred and more or less consistent
with our model. Suppose further that one wants to be able
to reject the null hypothesis (``an axisymmetric system'') with a specified
confidence level $\alpha$ (e.g. 95\%). To this end, any observed test parameter
has to lie outside the $1-\alpha$ region of the probability curve of the
axisymmetric model [2]+[3]. Since we assumed that the galaxy is compatible with
the barred model, we know a priori that this parameter obeys the distribution
[1]+[2]. Using this information, one can calculate the probability that the
parameter indeed lies outside this region.

\placefigure{fig_probdep}

In fig. \ref{fig_probdep}, the probabilities for rejection of the null hypothesis are
shown for several confidence levels, as a function of the strength of the bar (a value
of
$1$ corresponds to the model described in section \ref{modeldef}). These results are
calculated using Gaussian fits to the distributions of the test parameters. The 
estimates for bar strengths other than $1$ are based on a linear interpolation of the
parameter distributions between the axisymmetric and the barred model. This
approximation
is justified by the fact that our model is to a high degree linear, and that the test
parameters of both extreme models have more or less the same distribution.  Of course,
one
should realize that these numbers are based on a limited number Monte-Carlo
simulations,
and these values should therefore be considered only as rough estimates.

The third test parameter clearly turns out to be the most discriminating one.
Presumably, this is a consequence of the fact that the deviation of the proper motions
and the radial velocities happen to work in the same direction. For samples containing
700 stars, it is the only test that offers a reasonable chance to rule out an
axisymmetric model at a high level of confidence. The first test turns out to be less
sensitive, but still offers a reasonable discriminating power for samples containing
(at
least) 1400 stars. The second test, which only involves proper motions, has the
poorest
score.

These probability numbers indicate the a priori chance
that a particular type of observations will lead to positive results. In this way,
they are very useful for the estimation of the required sample size.

\section{Conclusions}

Although many observations point in the direction of a triaxial structure in
the central region of our Galaxy, there has been so far very little evidence
for this in the kinematical properties of the stars. In section \ref{sect_lv},
we have shown that, to a large degree, this can be explained by the presence of
a velocity dispersion in the motion of the stars. Such a dispersion diminishes
the effects of the bar on the distribution function, by ``smearing them out''
over a large portion of phase space. As a consequence, it turns out to be
very hard to prove triaxiality from an $l$-$v$ diagram alone, even if a
very large number of stars is involved, unless one has some additional 
information concerning the potential, obtained in an other way.

In the remainder of the article, we have shown that if distances and/or proper
motions of the stars are known, it is possible to construct test parameters
which discriminate between axisymmetric and triaxial models. Because distances
and proper motions are usually subject to large errors, the tests were designed
to be ``robust'' with respect to these values.  Moreover, the use of only first
order moments has a distinct advantage over second order moments (on which e.g.
vertex deviation is based), because higher order moments are less well
constrained by the data. Monte-Carlo simulations were further used in order to
estimate the probability distribution function of these test parameters for
different models. It turns out that the most powerful test incorporates a
combination of radial velocities and proper motions. However, even in this
case, one needs a large sample size (of the order of 1000 stars) in order to
have a good chance to be able to rule out an axisymmetric model with a high
degree of confidence.

{}

\clearpage

\figcaption[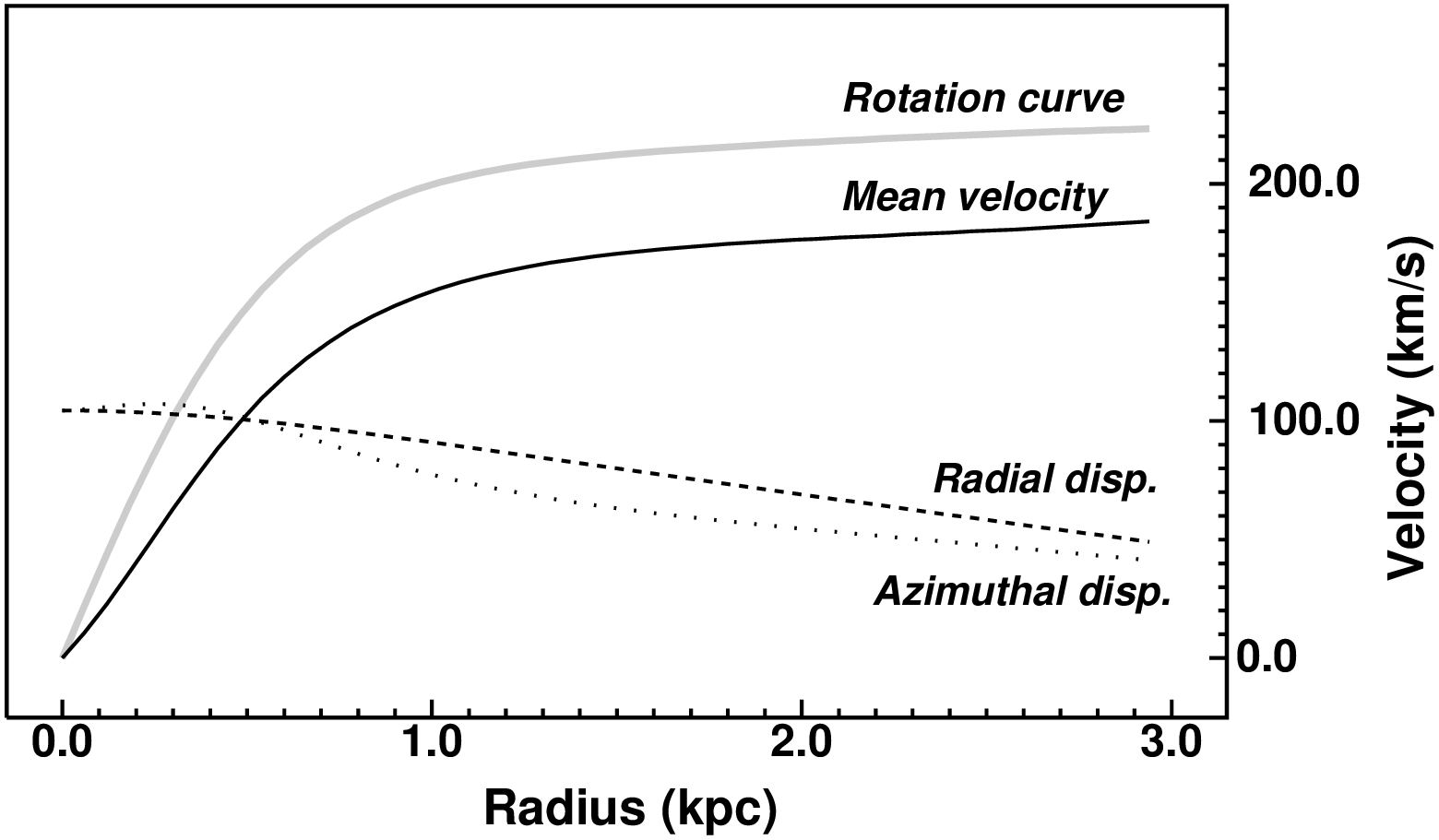]{
Rotation curve, mean velocity and dispersions of the axisymmetric
model.
\label{fig_axi} }

\figcaption[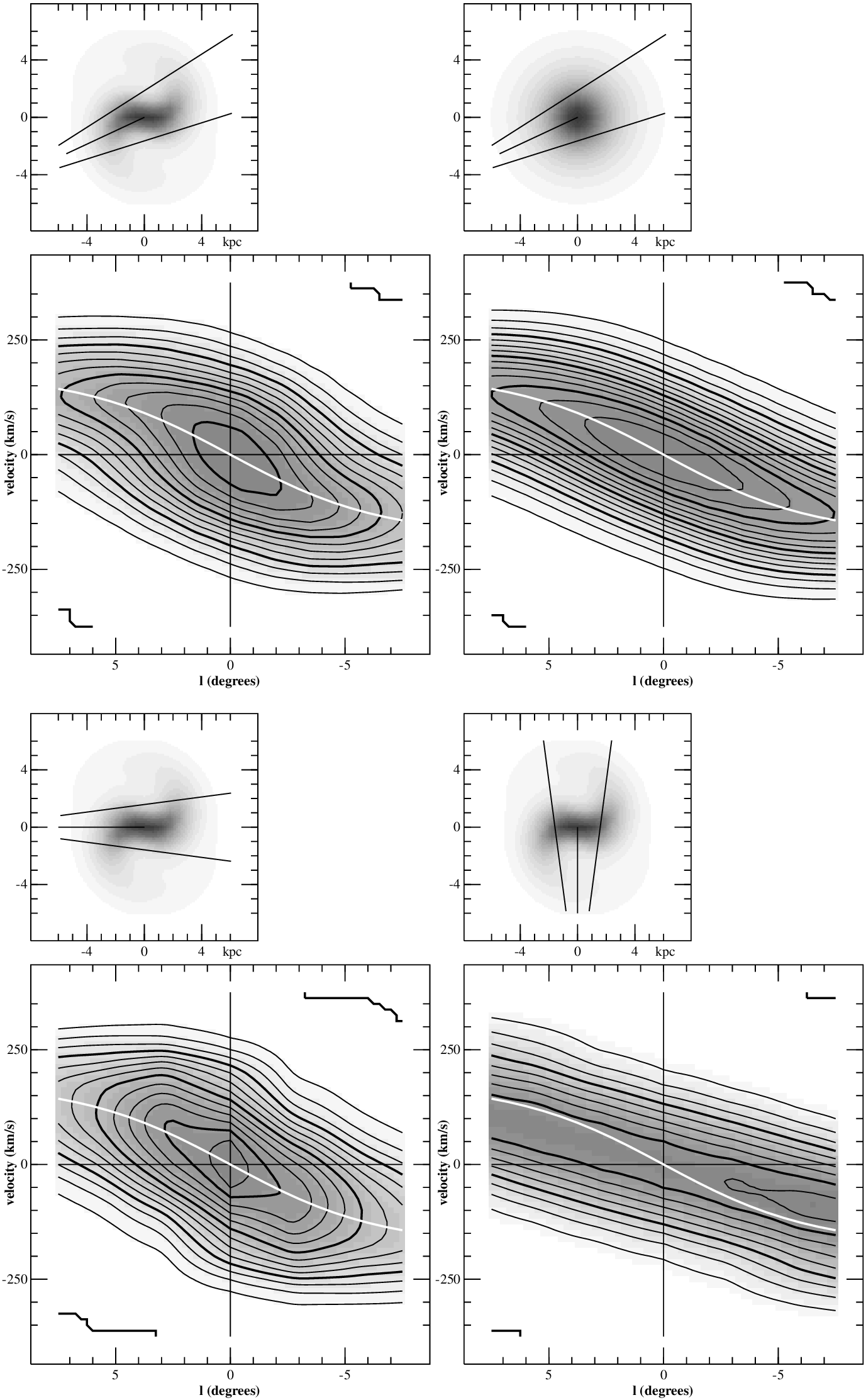]{
$l$--$v$ diagrams of the barred model (top left)
and the unperturbed system (top right). The axisymmetric rotation curve is
indicated
by the white line. The mass distribution and the orientation of the observer of
the system is shown in the small pictures. The bottom row shows the same bar with
different orientations.
\label{fig_lv} }

\figcaption[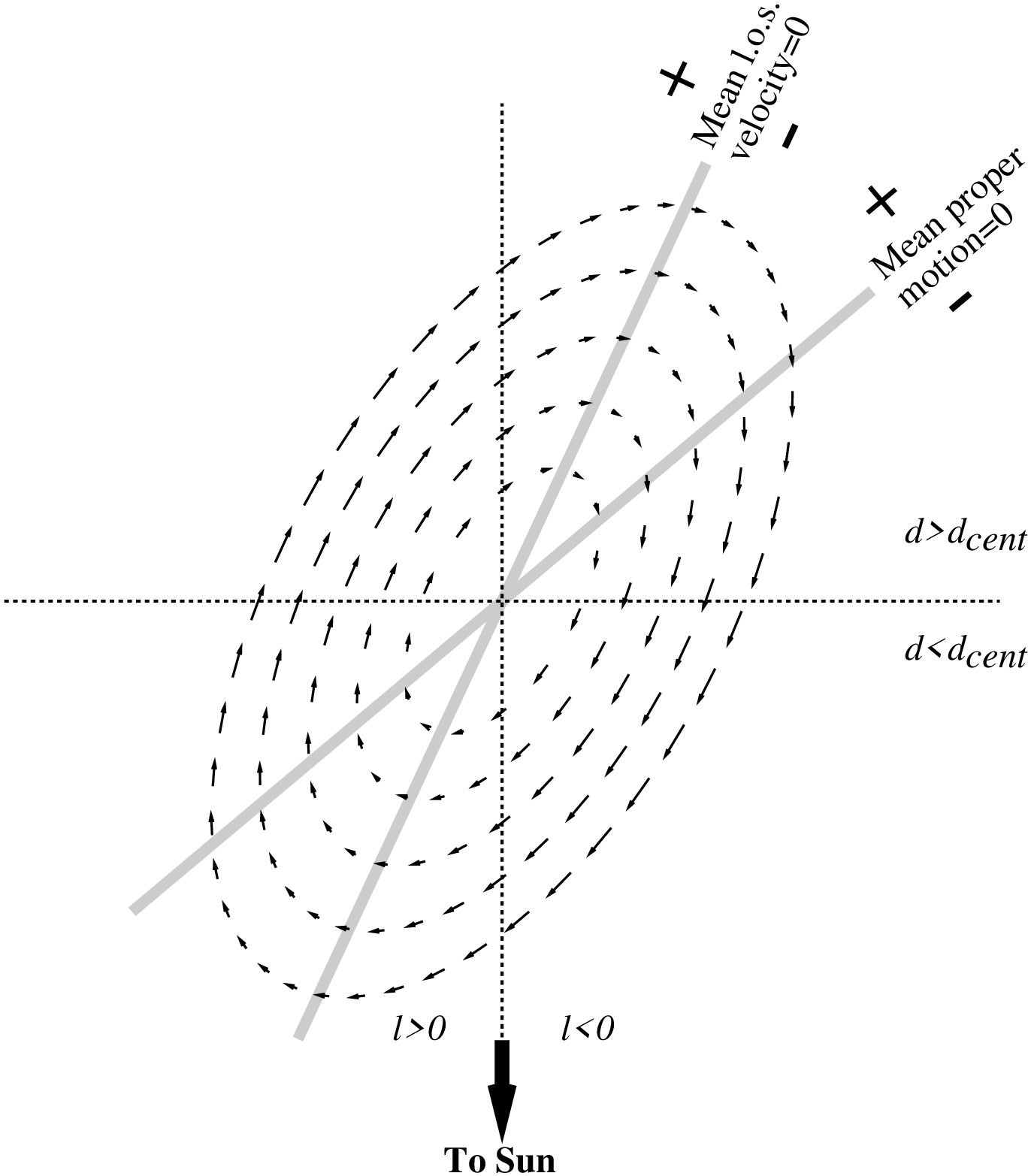]{
Schematic view of the streaming velocities in a barred stellar system. Note the
counterrotating streamlines that are present in the quadrants containing the
major axis. This is in contrast to axisymmetric systems, where counterrotating
stars are present only because of velocity dispersion.
\label{fig_sceme}}

\figcaption[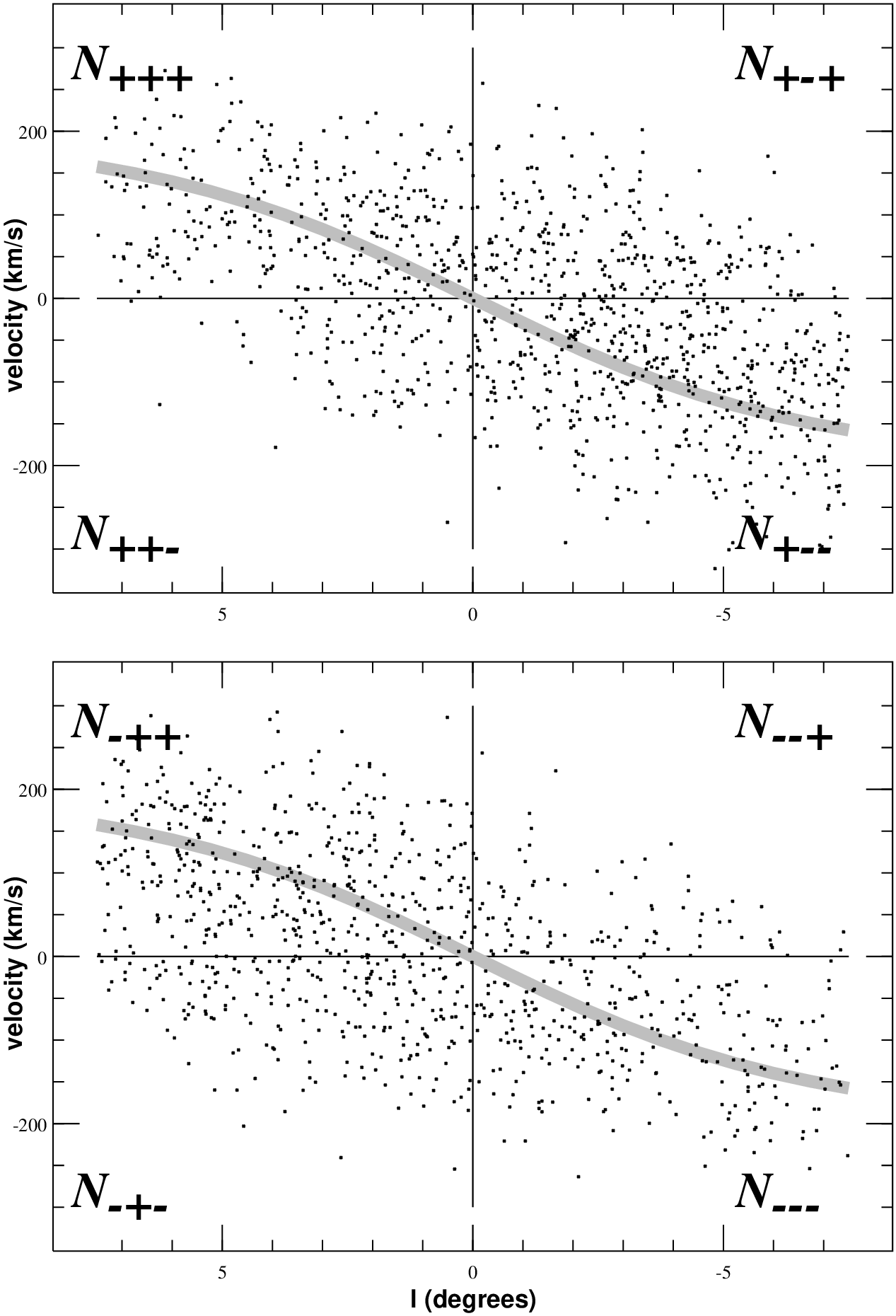]{
The $l$-$v$ diagrams of a simulated star sample, drawn from the barred 
model, for stars further away than the galactic centre (top panel),
and stars closer to the sun (bottom panel). 
The gray curves represent the axisymmetric rotation curve, and in each quadrant
of the graphs, the corresponding subsample index is marked (see text). The star
sample contains 700 stars.
\label{fig_distlv} }

\figcaption[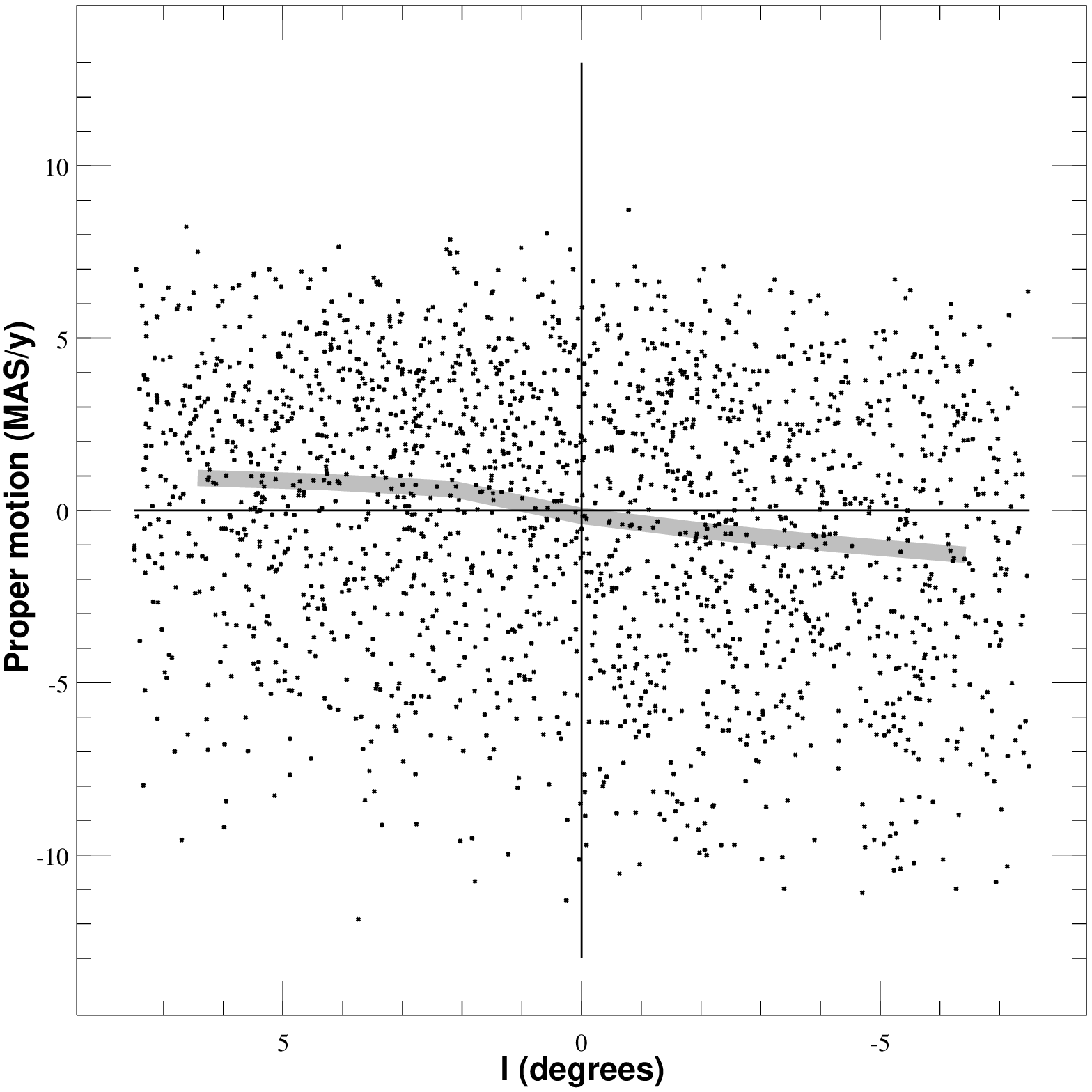]{
Proper motions of a simulated star sample (drawn from the barred model)
as a function of the galactic length. The gray line indicates the
averaged values.
\label{fig_prop0} }

\figcaption[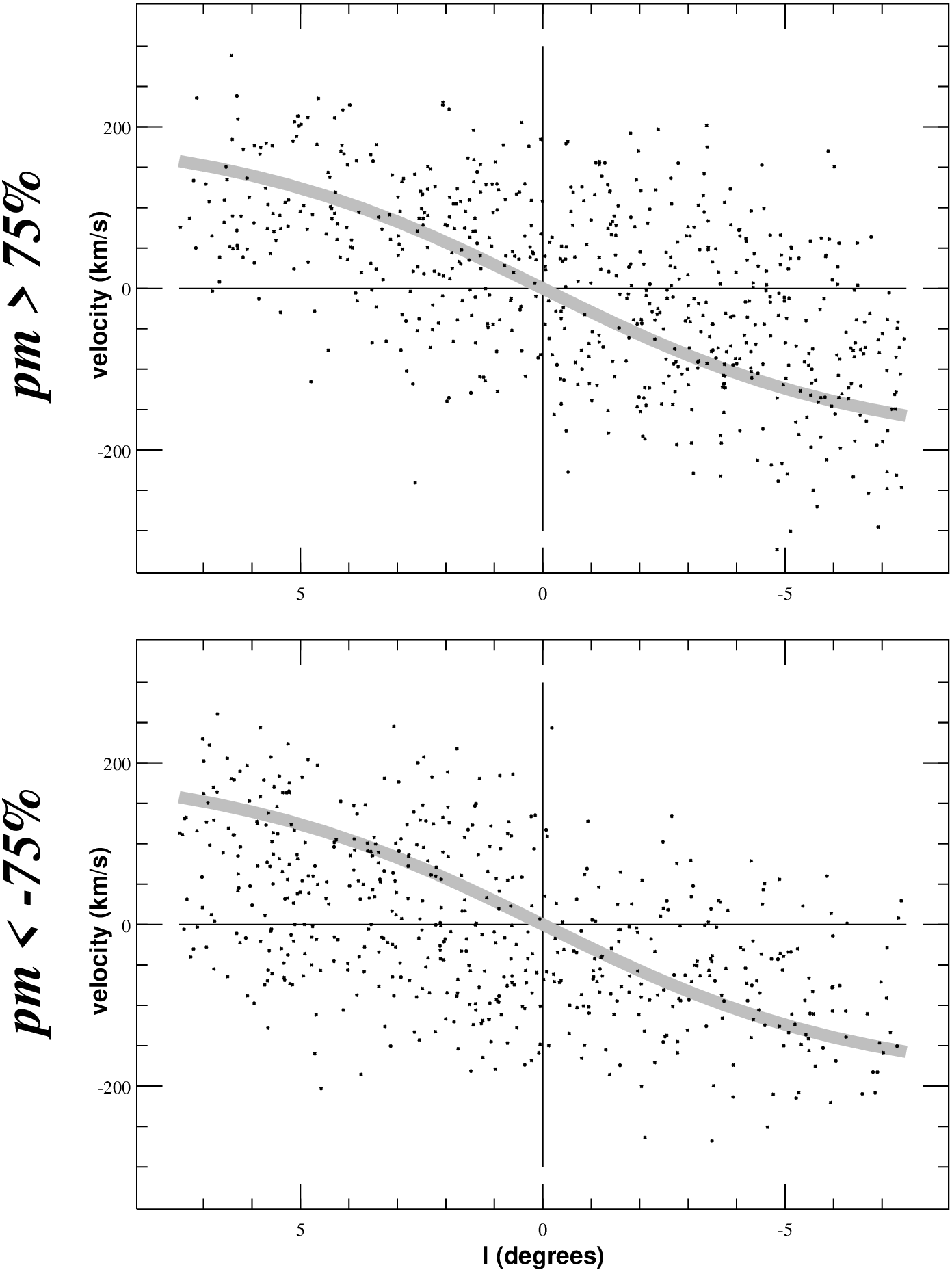]{
The $l$-$v$ diagrams of a simulated star sample
for the upper 75\% subset (top panel) and the lower 25\% (bottom panel).
The sample contains 700 stars, and is drawn from the barred model. The gray
lines indicate the axisymmetric rotation curve.
\label{fig_prop} }

\figcaption[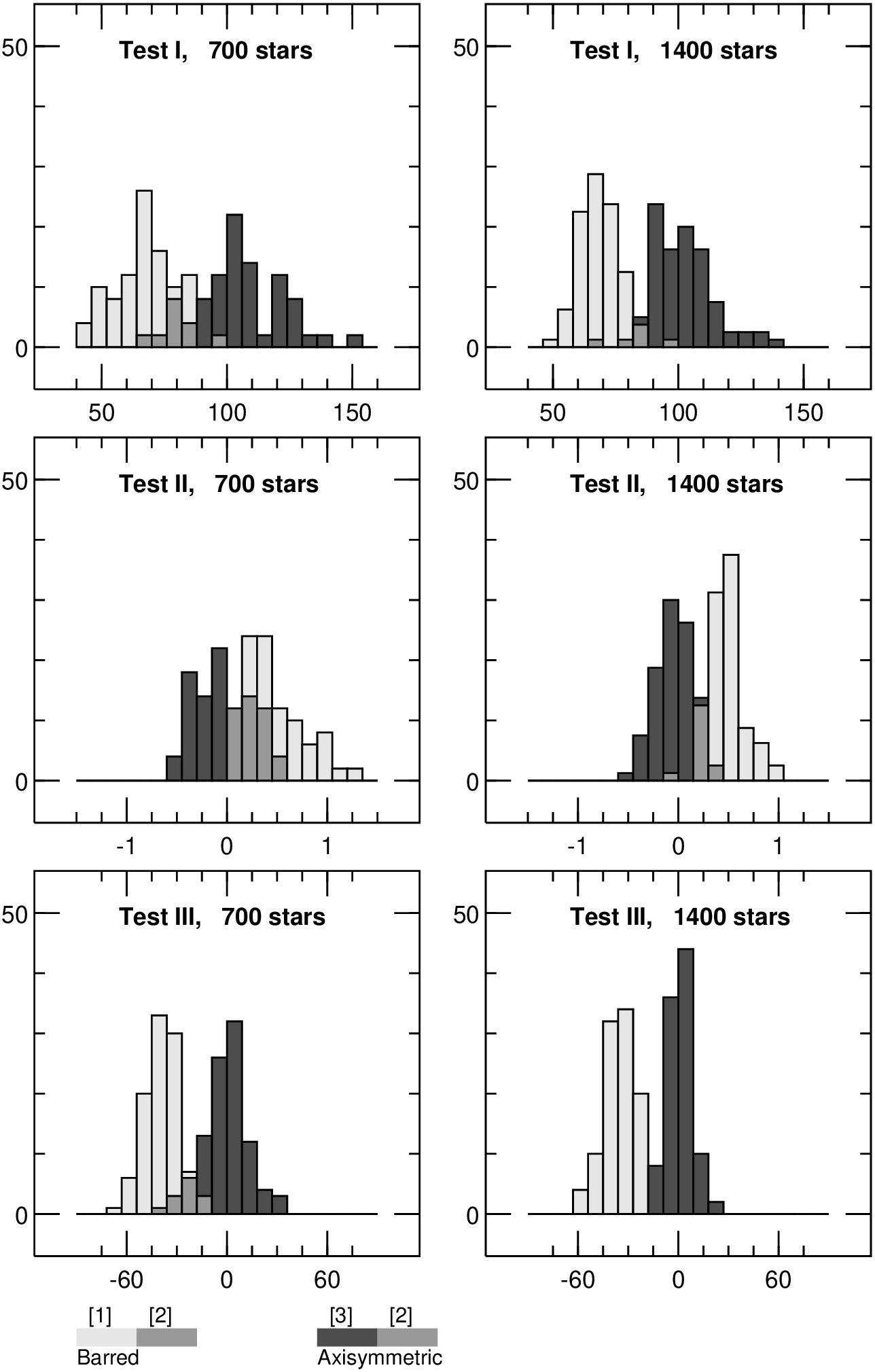]{
Distribution functions of the test parameters for different models.
\label{fig_pardst} }

\figcaption[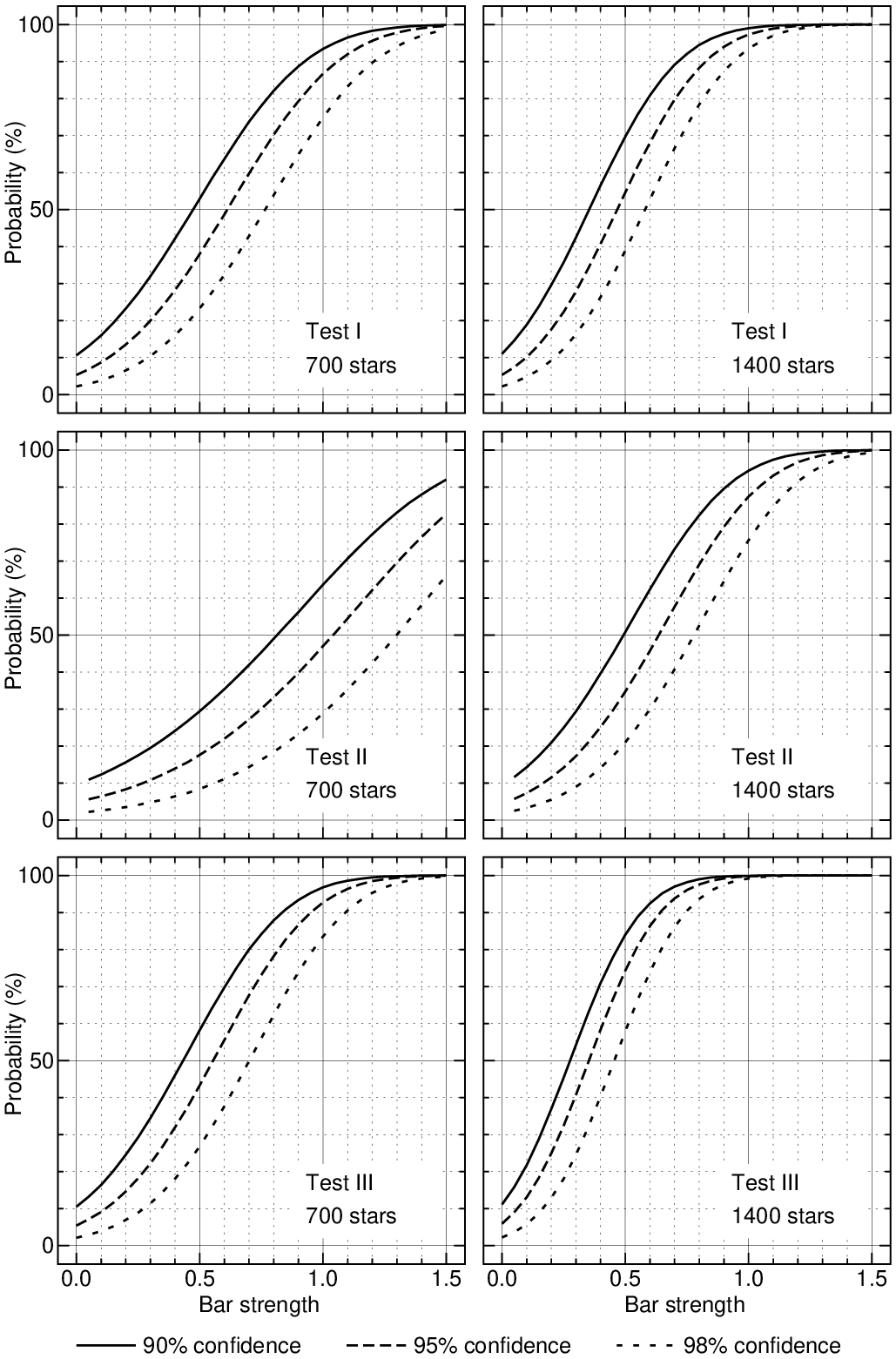]{
Probabilities for rejection of the axisymmetric hypothesis
for the three test parameters, at different levels of confidence, and
for various bar strengths (expressed as a fraction of the strength
of the bar in our model).
\label{fig_probdep} }

\begin{figure*}
\plotone{f1.ps}
\end{figure*}

\begin{figure*}
\plotone{f2.ps}
\end{figure*}

\begin{figure*}
\plotone{f3.ps}
\end{figure*}

\begin{figure*}
\plotone{f4.ps}
\end{figure*}

\begin{figure*}
\plotone{f5.ps}
\end{figure*}

\begin{figure*}
\plotone{f6.ps}
\end{figure*}

\begin{figure*}
\plotone{f7.ps}
\end{figure*}

\begin{figure*}
\plotone{f8.ps}
\end{figure*}

\end{document}